\documentclass[twocolumn,amssymb,amssymb,aps,pra]{revtex4} 
\usepackage{epsfig}
\begin{document}

\title{On the stability of general relativistic geometric thin disks}

\author{Maximiliano Ujevic}\email[e-mail: ]{mujevic@ime.unicamp.br}

\author{Patricio S. Letelier}\email[e-mail: ]{letelier@ime.unicamp.br}
\affiliation{Departamento de Matem\'atica Aplicada, Instituto de
Matem\'atica, Estat\'{\i}stica e Computa\c{c}\~ao Cient\'{\i}fica \\
Universidade Estadual de Campinas, 13081-970, Campinas, SP, Brasil}

\begin{abstract}
The stability of general relativistic thin disks is investigated under a
general first order perturbation of the energy momentum tensor.  In
particular, we consider temporal, radial and azimuthal ``test matter"  
perturbations of the quantities involved on the plane $z=0$. We study the
thin disks generated by applying the ``displace, cut and reflect" method,
usually known as the image method, to the Schwarzschild metric in
isotropic coordinates and to the Chazy-Curzon metric and the
Zipoy-Voorhees metric ($\gamma$-metric) in Weyl coordinates. In the case
of the isotropic Schwarzschild thin disk, where a radial pressure is
present to support the gravitational attraction, the disk is stable and
the perturbation favors the formation of rings. Also, we found the
expected result that the thin disk models generated by the Chazy-Curzon
and Zipoy-Voorhees metric with only azimuthal pressure are not stable
under a general first order perturbation.

PACS: 04.40.Dg,  98.62.Hr, 04.20.Jb
\end{abstract}

\maketitle

\section{Introduction} \label{section1}

In the last decades, analytical axially symmetric disk solutions have
appeared in both Newtonian and General Relativistic formulations. In the
context of General Relativity, several exact solutions were found, among
them the static disks without radial pressure studied by Bonnor and
Sackfield \cite{bon:sac} and Morgan and Morgan \cite{mor:mor1}. Disks with
radial pressure and with radial tension have been considered by Morgan and
Morgan \cite{mor:mor2} and Gonz\'ales and Letelier \cite{gon:let1},
respectively. Also, stationary disk models including electric fields
\cite{led:zof}, magnetic fields \cite{let}, and both electric and magnetic
fields \cite{kat:bic} have been studied. Self similar static disks were
considered by Lynden-Bell and Pineault \cite{lyn:pin} and Lemos
\cite{lem}. Furthermore, the superposition of black holes with static
disks were analyzed by Lemos and Letelier
\cite{lem:let1,lem:let2,lem:let3} and Klein \cite{kle}. Bi\v{c}\'ak and
Ledvinka \cite{bic:led} found relativistic counter-rotating thin disks as
sources of Kerr type metrics, and Bi\v{c}\'ak, Lynden-Bell and Katz
\cite{bic:lyn} obtained static disks as sources of known vacuum spacetimes
from the Chazy-Curzon metric \cite{cha,cur} and Zipoy-Voorhees metric
\cite{zip,voo}, also Bi\v{c}\'ak, Lynden-Bell and Pichon \cite{bic:lyn2}
found an infinite number of new static solutions. Recently, exact
solutions for thin disks with single and composite halos of matter
\cite{vog:let}, thin disks made of charged dust \cite{vog:let2}, thin
disks made of charged perfect fluid \cite{vog:let3}, and thick disks
\cite{gon:let2} were obtained. For a survey on relativistic gravitating
disks, see \cite{sem}.

It is well known that the formation of stellar systems lays in its
stability. The study of stability, analytically or numerically, is vital
to the acceptance and applicability of the different models found in the
literature to describe stellar structures found in Nature. On the other
hand, the study of different types of perturbations, when applied to
stellar structures, usually give an insight on the formation of bars,
rings or different stellar patterns. Also, a perturbation may cause the
collapse of an stable object with the posterior appearance of a different
kind of structure.

In Newtonian theory, perturbations have been made extensively through the
years in analytical calculations and numerical experiments. An analytical
treatment of the stability of thin disks can be found in Refs.  
\cite{bin:tre,fri:pol} and references therein. In General Relativity, the
stability analysis is usually done studying the particle motion along
geodesics and not perturbing the energy momentum tensor of the fluid and
its conservation equations. The stability of particle motion along
geodesics have been studied transforming the Rayleigh criteria of
stability \cite{ray,lan:lif} into a General Relativistic formulation, see
\cite{let2} and reference therein. By using this criterion, the stability
of orbits around black holes surrounded by disks, rings and multipolar
fields were analyzed in \cite{let2}. Also, it was employed in
\cite{vog:let} to test stability in the isotropic Schwarzschild thin disk,
and thin disks of single and composite halos generated from the Buchdahl
solution \cite{buc} and Narlikar, Patwardhan and Vaidya solution
\cite{nar:pat} by using the displace, cut and reflect method. For the
stability of circular orbits in stationary axisymmetric spacetimes see
Ref. \cite{bar,abr:pra}. The stability of circular orbits of the
Lemos-Letelier solution \cite{lem:let2} for the superposition of a black
hole and a flat ring, as well as other aspects of this solution, are
considered in detail in \cite{sem:zac,sem:zac2,sem2}.  Bi\v{c}\'ak,
Lynden-Bell and Katz \cite{bic:lyn} analyzed the stability of the
Chazy-Curzon and Zipoy-Voorhees thin disks without radial pressure or
tension studying their velocity curves and specific angular momentum,
finding that they are not stable for highly relativistic disks. A general
stability study of a relativistic fluid, with both bulk and dynamical
viscosity, perturbing its energy momentum tensor was done by Seguin
\cite{seg}. However, he consider the coefficients of the perturbed
variables as constants, i.e., local perturbations. Fact that in general is
too restrictive.

The purpose of this work is to study numerically the stability of several
analytical thin disk solutions in the context of General Relativity. The
stability is studied performing a general first order perturbation in the
 temporal, radial and azimuthal components of the quantities involved
in the  energy momentum tensor of the fluid and analyzing the
corresponding perturbed conservation equations of motion.  The
perturbations considered do not modified the background metric obtained
from the solution of Einstein equations, i.e.they  are treated as ``test
matter".

The article is organized as follows. In Section II-A we present the
perturbed conservation equations for the general case, when the energy
momentum tensor is diagonal with the energy density and the three
pressures or tensions all different from zero. Later, in Section II-B, we
specialize these perturbed equations for the thin disk case where we have
a distributed energy momentum tensor with support on the plane $z=0$. In
Section III, we analyze the perturbed conservation equations to study the
stability of the isotropic Schwarzschild thin disk, which contains radial
and azimuthal pressures. In Section IV, we analyze the perturbed
conservation equations to study the stability of the Chazy-Curzon and
Zipoy-Voorhees thin disks { with only azimuthal pressure. It is worth
to mention that  different Chazy-Curzon and Zipoy-Voorhees
thin disks models with radial tensions or pressures can be built by using a
different set of coordinates, but these models do not usually satisfy all the
physical requirements, see \cite{gon:let1}.} Finally, in Section V, we
summarize and discuss the presented results.

\section{Perturbed equations} \label{section2}

\subsection{General case}

Let us consider the general static, axially symmetric metric

\begin{equation}
ds^2= - e^{2 \Psi_1} dt^2 + e^{2 \Psi_2} r^2 d\varphi^2 +
e^{2 \Psi_3} (dr^2+dz^2), \label{metric}
\end{equation}

\noindent where $\Psi_1$, $\Psi_2$ and $\Psi_3$ are functions of the
variables $(r,z)$. (Our conventions are:  $G=c=1$.  
Metric signature +2. Partial and covariant derivatives with respect to the
coordinate $x^\mu$ denoted by $,\mu$ and $;\mu$, respectively.)

The energy momentum tensor of the fluid $T^{\mu\nu}$ satisfies Einstein
field equations $G_{\mu\nu} = \kappa T_{\mu\nu}$, and, when written in its
rest frame, is diagonal with components ($-\rho,p_r,p_\varphi,p_z$), where
$\rho$ is the energy density and ($p_r,p_\varphi,p_z$) are the radial,
azimuthal and axial pressures or tensions respectively. With these
definitions, the energy momentum tensor can be written as

\begin{equation}
T^{\mu\nu} = \rho U^\mu U^\nu + p_r X^\mu X^\nu + p_\varphi Y^\mu Y^\nu + 
p_z Z^\mu Z^\nu, \label{tmunu}
\end{equation}

\noindent where $U^\mu$, $X^\mu$, $Y^\mu$ and $Z^\mu$ are the 4-vectors of
the orthonormal tetrad

\begin{eqnarray}
&&U^\mu = e^{-\Psi_1} (1,0,0,0), \nonumber \\
&&X^\mu = e^{-\Psi_3} (0,1,0,0), \nonumber \\
&&Y^\mu = \frac{e^{-\Psi_2}}{r} (0,0,1,0), \nonumber \\
&&Z^\mu = e^{-\Psi_3} (0,0,0,1), \label{tetrad}
\end{eqnarray}

\noindent which satisfy,

\begin{eqnarray}
&&-U^\mu U_\mu = X^\mu X_\mu = Y^\mu Y_\mu = Z^\mu Z_\mu = 1, \nonumber \\
&&U^\mu X_\mu = U^\mu Y_\mu = U^\mu Z_\mu = 0 \nonumber \\
&&X^\mu Y_\mu = X^\mu Z_\mu = Y^\mu Z_\mu = 0. \label{ortho}
\end{eqnarray}

\noindent The quantity $U^\mu$ is the time-like four-velocity of the fluid
and the quantities $X^\mu$, $Y^\mu$ and $Z^\mu$ are the space-like
principal directions of the fluid.

In the axially symmetric case, all the quantities involved in the energy
momentum tensor are functions of the coordinates ($r,z$) only, and so are
the coefficients of the perturbed conservation equations. With this in
mind, we construct the general perturbation $A^\mu_P$ of a quantity
$A^\mu$ in the form

\begin{equation}
A^\mu_P(t,r,\varphi,z) = A^\mu(r,z) + \delta A^\mu(t,r,\varphi,z), 
\label{pertur}
\end{equation}

\noindent where $A^\mu(r,z)$ is the unperturbed quantity and $\delta
A^\mu(t,r,\varphi,z)$ is the perturbation. Applying the perturbation
(\ref{pertur})  in (\ref{tmunu}) we obtain that

\begin{equation}
T^{\mu\nu}_P(t,r,\varphi,z) = T^{\mu\nu}(r,z) + \delta 
T^{\mu\nu}(t,r,\varphi,z).
\end{equation}

\noindent Henceforth, we assume that the perturbed energy momentum tensor
does not modify the background metric found by solving the Einstein field
equations $G_{\mu\nu} = \kappa T_{\mu\nu}$, i.e. the perturbation $\delta
T^{\mu\nu}$ acts as a test fluid. With this assumption, the perturbed
energy momentum equations for the problem can be written as

\begin{equation}
(\delta T^{\mu\nu})_{;\nu} = 0 \label{ptmunu}.
\end{equation}

\noindent Writing in explicit form the four equations from (\ref{ptmunu})
and discarding terms of order greater or equal to $(\delta)^2$, we obtain
the perturbed conservation equations for the system

\begin{widetext}
$\mu=t$
\begin{eqnarray}
&&\delta U^t_{,t} (\rho U^t) + \delta U^\alpha_{,\alpha} (\rho 
U^t) + \delta X^t_{,r} (p_r X^r) + \delta Y^t_{,\varphi} (p_\varphi 
Y^\varphi) + \delta Z^t_{,z} (p_z Z^z) + \delta \rho_{,t} (U^t U^t) 
\nonumber \\
&&+ \delta U^r {\rm F}(t,r,\rho U^t) + \delta U^z {\rm F}(t,z,\rho U^t) +
\delta X^t {\rm F}(t,r, p_r X^r) + \delta Z^t {\rm F}(t,z,p_z Z^z) = 0,
\label{eqgeralt}
\end{eqnarray} 

$\mu=r$
\begin{eqnarray}
&&\delta U^r_{,t} (\rho U^t) + \delta X^r_{,r} (p_r X^r) + \delta 
X^\alpha_{,\alpha} (p_r X^r) + \delta Y^r_{,\varphi} (p_\varphi  
Y^\varphi) + \delta Z^r_{,z} (p_z Z^z) +\delta p_{r,r} (X^r X^r)  + 2 
\delta U^t (\rho U^t \Gamma^r_{tt}) \nonumber \\
&&+ 2 \delta X^r {\rm G}(r,r,p_r X^r) + \delta X^z 
{\rm F}(r,z, p_r X^r) + 2 \delta Y^\varphi (p_\varphi Y^\varphi 
\Gamma^r_{\varphi\varphi}) + \delta Z^r {\rm F}(r,z,p_z Z^z) + 2 \delta 
Z^z (p_z Z^z \Gamma^r_{zz}) \nonumber \\
&&+ \delta \rho (U^t U^t \Gamma^r_{tt}) + \delta p_r {\rm G}(r,r,X^r X^r) 
+ \delta p_\varphi (Y^\varphi Y^\varphi \Gamma^r_{\varphi\varphi} ) + 
\delta p_z (Z^z Z^z \Gamma^r_{zz}) = 0,
\label{eqgeralr}
\end{eqnarray} 

$\mu=\varphi$
\begin{eqnarray}
&&\delta U^\varphi_{,t} (\rho U^t) + \delta X^\varphi_{,r} (p_r X^r) +
\delta Y^\varphi_{,\varphi} (p_\varphi Y^\varphi) + \delta 
Y^\alpha_{,\alpha} (p_\varphi Y^\varphi) + \delta Z^\varphi_{,z}
(p_z Z^z) + \delta p_{\varphi,\varphi} (Y^\varphi Y^\varphi) \nonumber \\
&&+ \delta X^\varphi {\rm F}(\varphi,r,p_r X^r) + \delta Y^r {\rm
F}(\varphi,r,p_\varphi Y^\varphi) + \delta Y^z {\rm F}(\varphi,z,
p_\varphi Y^\varphi) + \delta Z^\varphi {\rm F}(\varphi,z,p_z Z^z) = 0,
\label{eqgeralvarphi}
\end{eqnarray} 

$\mu=z$
\begin{eqnarray}
&&\delta U^z_{,t} ( \rho U^t) + \delta X^z_{,r} (p_r X^r) + \delta 
Y^z_{,\varphi} (p_\varphi Y^\varphi) + \delta Z^z_{,z} (p_z Z^z) + \delta 
Z^\alpha_{,\alpha} (p_z Z^z) + \delta p_{z,z} (Z^z Z^z) + 2 \delta U^t 
(\rho U^t \Gamma^z_{tt}) \nonumber \\
&&+ 2 \delta X^r (p_r X^r \Gamma^z_{rr}) + \delta X^z {\rm F}(z,r,p_r 
X^r) + 2 \delta Y^\varphi (p_\varphi Y^\varphi \Gamma^z_{\varphi\varphi}) 
+ \delta Z^r {\rm F}(z,r,p_z Z^z) + 2 \delta Z^z {\rm G}(z,z,p_z Z^z) 
\nonumber \\
&&+ \delta \rho (U^t U^t \Gamma^z_{tt}) + \delta p_r (X^r X^r 
\Gamma^z_{rr}) + \delta p_\varphi (Y^\varphi Y^\varphi \Gamma^z_{\varphi 
\varphi}) + \delta p_z {\rm G}(z,z,Z^z Z^z) = 0, \label{eqgeralz}
\end{eqnarray} 

\noindent where

\begin{eqnarray}
&&{\rm F}(I,J,K) = K_{,J} + K (2 \Gamma^I_{IJ} + \Gamma^\alpha_{\alpha
J}), \\
&&{\rm G}(I,J,K) = K_{,J} + K (\Gamma^I_{IJ} + \Gamma^\alpha_{\alpha J}),
\end{eqnarray}
\end{widetext}

\noindent and $\Gamma^\alpha_{\beta\gamma}$ are the Christoffel symbols.

We want the perturbed tetrad to remain orthonormal. Therefore, to
guarantee the orthonormal conditions (\ref{ortho}) we obtain the following
relations,

\begin{eqnarray}
&&\delta U^t = \delta X^r = \delta Y^\varphi = \delta Z^z = 0, \nonumber 
\\
&&\delta X^t = \xi_1 \delta U^r, \hspace{1.08cm} \xi_1 = -X_r/U_t, 
\nonumber \\
&&\delta Y^t = \xi_2 \delta U^\varphi, \hspace{1.06cm} \xi_2 = 
-Y_\varphi/U_t, \nonumber \\
&&\delta Z^t = \xi_3 \delta U^z, \hspace{1.1cm} \xi_3 = -Z_z/U_t, 
\nonumber \\
&&\delta X^\varphi = \xi_4 \delta Y^r, \hspace{0.98cm} \xi_4 = 
-X_r/Y_\varphi, \nonumber \\
&&\delta X^z = \xi_5 \delta Z^r, \hspace{1.04cm} \xi_5 = -X_r/Z_z, 
\nonumber \\
&&\delta Y^z = \xi_6 \delta Z^\varphi, \hspace{1.03cm} \xi_6 = 
-Y_\varphi/Z_z. \label{conditions1}
\end{eqnarray}

\noindent It is clear from Eqs. (\ref{conditions1}) that in order to have
a consistent perturbation model, the tetrad ought to be perturbed. In our
case, we are interested in perturbations in the velocity components and
for that reason the $t$ component of the tetrad must be perturbed.

\subsection{Thin disk case}

Thin disks are characterized by a distribution valued energy momentum
tensor that is a delta function with support on the plane $z=0$. In this
case, the axial pressure $p_z$ is equal to zero and the energy density,
radial and azimuthal pressures or tensions are only functions of the
radial coordinate $r$. So, the energy momentum tensor can be cast into the
form

\begin{equation}
T^{\mu\nu} = (\sigma U^\mu U^\nu + p_r X^\mu X^\nu + p_\varphi Y^\mu 
Y^\nu) \delta (z). \label{thintmunu}
\end{equation}

\noindent where $\sigma$ is the surface energy density and
($U^\mu,X^\mu,Y^\mu$) are the components of the tetrad previously defined.
Note that when we define the energy momentum tensor as above, the fluid
quantities ($\sigma,p_r,p_\varphi$) are not the ``true" fluid quantities
($\tilde{\sigma}, \tilde{p_r}, \tilde{p_\varphi}$). These two sets of
quantities are related by

\begin{equation}
\tilde{\sigma} = \sqrt{g_{zz}} \sigma, \hspace{0.3cm} \tilde{p_r} = 
\sqrt{g_{zz}} p_r, \hspace{0.3cm} \tilde{p_\varphi} = \sqrt{g_{zz}} 
p_\varphi. \label{real}
\end{equation}

Using the perturbed conservation equations (\ref{ptmunu}) and the
distributed energy momentum tensor (\ref{thintmunu}) we find that

\begin{equation}
(\delta T^{\mu\nu})_{;\nu} \delta(z) + \delta T^{\mu\nu} 
[\delta(z)]_{,\nu} = 
0. \label{thinptmunugeral}
\end{equation}

\noindent In this work, the thin disk metrics considered are found by
applying the well known displace, cut and reflect method
\cite{bic:lyn,vog:let}. For that reason, the metric components depend on
the radial coordinate $r$ and the absolute value of $z$. If we define the
derivative of the absolute value of $z$ in $z=0$ equal to zero
($|z|_{,z=0}=0$), and perform in (\ref{thinptmunugeral}) the integration
on the coordinate $z$ ($\int \sqrt{g_{zz}} {\rm d}z$), we obtain that the
second term is equal to zero and the perturbed conservation equations for
thin disks can be written as

\begin{equation}
(\delta T^{\mu\nu})_{;\nu} |_{z=0} = 0. \label{thinptmunu}
\end{equation}

In general, we are interested in perturbations of the four-velocity and
the thermodynamic quantities on the disk. Thus, we set the perturbations
on the $Z^\mu$ tetrad vector equal to zero. Also, the perturbation $\delta
X^\varphi$ is not related to the perturbation in the radial or azimuthal
velocities and we can set it equal to zero. Note that perturbations in the
tetrad are independent of the perturbations in the fluid quantities and
the assumptions made above do not change the equation of state of the
fluid. So, we have

\begin{equation}
\delta Z^t = \delta Z^r = \delta Z^\varphi = \delta X^\varphi = 0, 
\label{conditions2}
\end{equation}

\noindent which implies from (\ref{conditions1}) that

\begin{equation}
\delta U^z = \delta X^z = \delta Y^z = \delta Y^r = 0, 
\end{equation}

\noindent and the perturbed tetrad is still orthonormal.

Due to the form of the perturbed energy momentum equation for thin disks
(\ref{thintmunu}), the connection coefficients are only functions of the
radial coordinate. Therefore, all the coefficients on the perturbed
conservation equations depend only on $r$ and we construct a general
perturbation of the form

\begin{equation}
\delta A^\mu(t,r,\varphi) = \delta A^\mu(r) e^{i(k_\varphi 
\varphi - wt)}. \label{perturbation}
\end{equation}
\noindent Hereafter $\delta A^\mu \equiv \delta A^\mu(r)$. With the
perturbation (\ref{perturbation}) and conditions (\ref{conditions1}) and
(\ref{conditions2}), the perturbed conservation equations
(\ref{thinptmunu}) can be written in explicit form as

\begin{widetext}
$\mu=t$
\begin{eqnarray}
&&\delta U^r_{,r} (\sigma U^t + \xi_1 p_r X^r) + \delta 
U^r [ {\rm F}(t,r,\sigma U^t) + \xi_{1,r} p_r X^r + \xi_1 {\rm F}(t,r,p_r 
X^r)] \nonumber \\
&&+ \delta U^\varphi [ i k_\varphi (\sigma U^t + \xi_2 p_\varphi 
Y^\varphi) ] + \delta \sigma (-i w U^t U^t) = 0, \label{thint}
\end{eqnarray}

$\mu=r$
\begin{eqnarray}
&&\delta p_{r,r} (X^r X^r) + \delta U^r [-iw(\sigma U^t + 
\xi_1 p_r X^r)] + \delta \sigma (U^t U^t \Gamma^r_{tt} ) \nonumber 
\\
&&+ \delta p_r {\rm G}(r,r,X^r X^r) + \delta p_\varphi 
(Y^\varphi Y^\varphi \Gamma^r_{\varphi\varphi})= 0, \label{thinr}
\end{eqnarray}

$\mu=\varphi$
\begin{equation}
\delta U^\varphi [-w (\sigma U^t + \xi_2 p_\varphi Y^\varphi)] + 
\delta p_\varphi (k_\varphi Y^\varphi Y^\varphi) = 0.
\label{thinvarphi}
\end{equation}
\end{widetext}

\noindent where we set $\delta p_z = 0$ in order to maintain the state
equation of the fluid on the disk. The above equations are the basic
equations for thin disks stability and will be studied in the next
sections. The general equations (\ref{eqgeralt})-(\ref{eqgeralz}) are
presented for completeness and for future works on the stability of thick
disks and other static axially symmetric structures.

\begin{figure*}
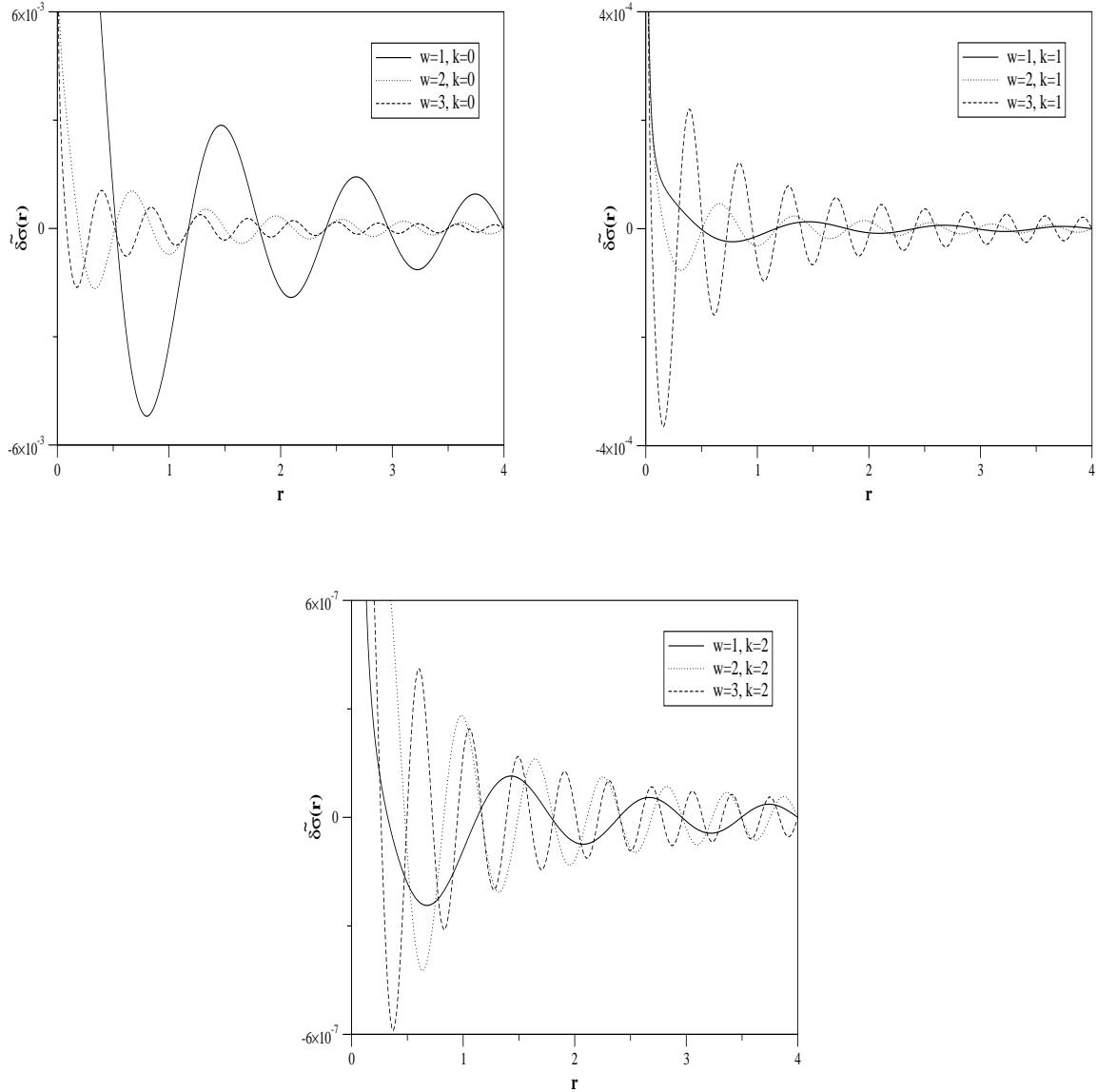

\epsfig{width=7cm, height=7cm, file=k0.eps} \hspace{1cm}
\epsfig{width=7cm, height=7cm, file=k1.eps}

\vspace{1.2cm} \epsfig{width=7cm, height=7cm, file=k2.eps}
\caption{Profiles of the energy density perturbation amplitudes 
of the isotropic Schwarzschild thin disk with parameters (a=0.5,m=0.4). 
The first three different $w$ modes for the first three wave number $k$ 
are plotted, we see that increasing the $w$ mode the number of 
oscillations within the disk increases while increasing the wave number 
$k$ the amplitude decreases.} 
\label{k0k1k2} 
\end{figure*}

\begin{figure*}
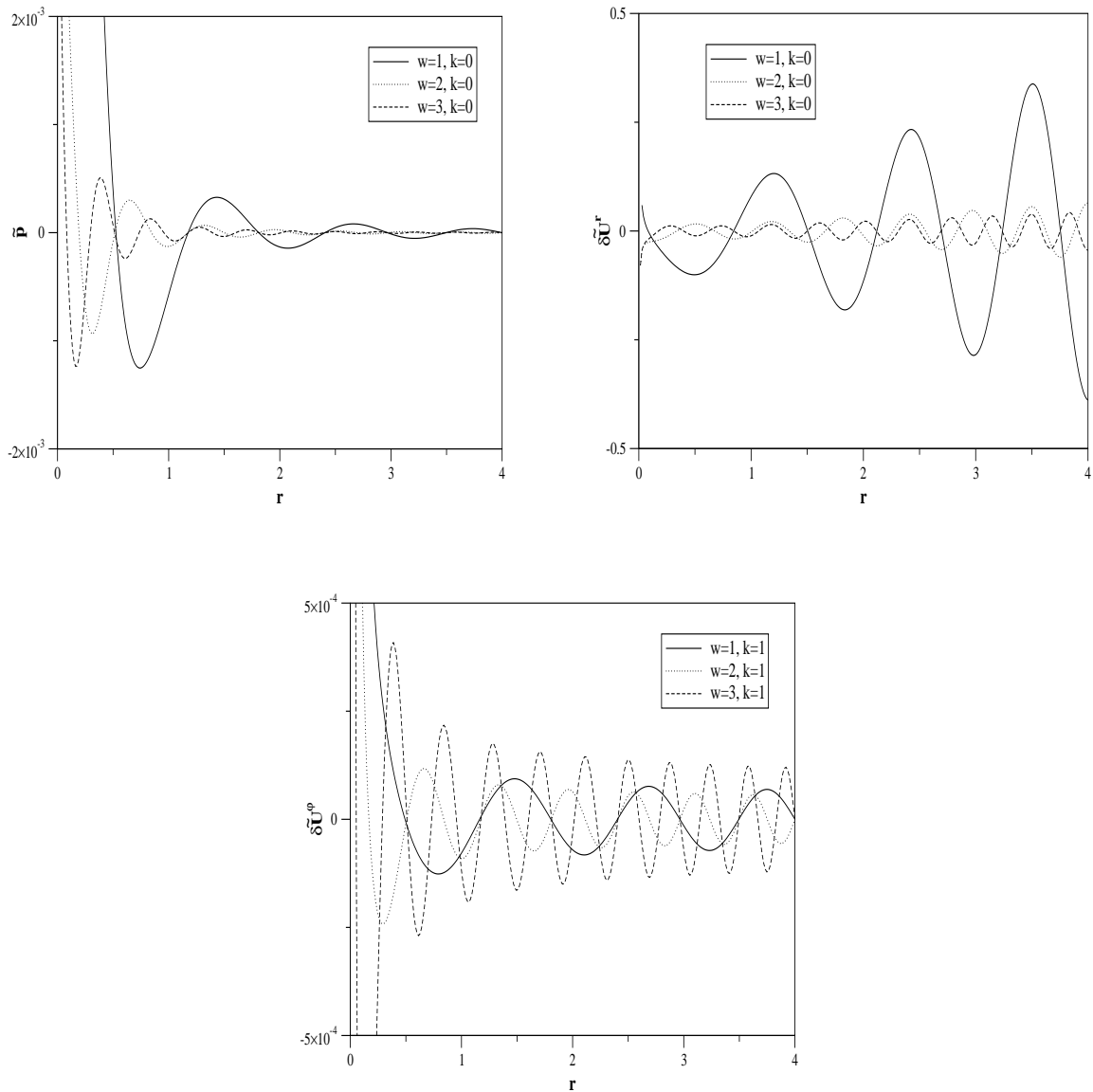

\epsfig{width=7cm, height=7cm, file=pressure.eps} \hspace{1cm}
\epsfig{width=7cm, height=7cm, file=veloradial.eps}

\vspace{1.2cm} \epsfig{width=7cm, height=7cm, file=velophi.eps}
\caption{Profiles of the true pressure perturbation, physical radial
velocity and physical azimuthal velocity amplitudes of the isotropic 
Schwarzschild thin disk with parameters (a=0.5,m=0.4). We see that the 
qualitative aspects of the pressure perturbation are the same of the 
energy density perturbation. The amplitude of the radial velocity 
increases when we get near the edge of the disk, in that circumstances we 
must compared these velocity values to the particle escape velocity to 
make the perturbation consistent with the model. The azimuthal 
velocity amplitude is almost constant far from the center of the disk.}
\label{velocity}
\end{figure*}

\section{Thin disks with radial and azimuthal pressures} \label{section4}

We start analyzing the isotropic Schwarzschild thin disk. This disk is
found by setting the metric functions ($\Psi_1,\Psi_2,\Psi_3$) in
(\ref{metric}) as

\begin{eqnarray}
&&\Psi_1 = \ln \left[ \frac{2 R - m}{2 R + m} \right], \nonumber \\
&&\Psi_2 = \Psi_3 = \ln \left[ 1 + \frac{m}{2 R} \right]^2, \label{psisch}
\end{eqnarray}

\noindent where $m$ is a positive constant and $R^2 = r^2 + (|z|+a)^2$,
and applying the displace, cut and reflect method to the resulting
isotropic Schwarzschild metric. With this operation, we find a thin disk
with surface energy density and equal pressures given by \cite{vog:let},

\begin{eqnarray}
&&\sigma= \frac{16 m a R_0^2}{\pi (m + 2 R_0)^5}, \label{sschwa}\\
&&p=p_r = p_\varphi = - \frac{8 m^2 a R_0^2}{\pi (m + 2 R_0)^5 
(m-2 R_0)}, \label{pschwa}
\end{eqnarray}

\noindent where $R_0$ means to evaluate $R$ at $z=0$. Recall that we are
not using the true fluid quantities, the quantities (\ref{sschwa}) and
(\ref{pschwa}) are related to the ones cited in \cite{vog:let} by Eqs.
(\ref{real}). From here and through the rest of the manuscript, all the
calculations are made in the variables ($\sigma, p_r, p_\varphi$) and the
final results presented in the figures expressed in terms of the true
physical variables.

We want our perturbations to be in accordance with the equation of state
of the fluid, i.e. $p = p(r)$ and $\sigma = \sigma(r)$.  Thus,
$\delta p_r$ and $\delta \sigma$ satisfy

\begin{eqnarray}
&&\delta p = \delta p_r = \delta p_\varphi = p_{,r} {\rm d}r, \\
&&\delta \sigma = \sigma_{,r} {\rm d}r,
\end{eqnarray}

\noindent from which we find the useful relation

\begin{equation}
\delta p = \left( \frac{ p_{,r}}{\sigma_{,r}}\right) \delta 
\sigma. \label{psigma}
\end{equation}

Substituting $\delta U^r$ and $\delta U^\varphi$ in Eq. (\ref{thint}) from
Eqs. (\ref{thinr}) and (\ref{thinvarphi}), and using relation
(\ref{psigma}), we find a second order differential equation for the
energy density perturbation $\delta \sigma$ of the form

\begin{equation}
A_S \delta \sigma_{,rr} + B_S \delta \sigma_{,r} + C_S \delta 
\sigma = 0, 
\label{2order}
\end{equation}

\noindent where ($A_S,B_S,C_S$) are functions of ($r,a,m,w,k_\varphi$),
see Appendix A.

Note that the exact thin disk metrics we are considering are infinite in
the radial direction. So, in order to study the stability of the disks we
need a criterion to make a cut-off in the radial coordinate to create a
finite disk. From \cite{vog:let} we see that the energy density of the
disks decreases rapidly enough in principle to define a cut-off radius.  
The cut-off radius $R_{cut}$ of the disk is set by the following
criterion: the matter within the thin disk formed by the cut-off radius is
more than 90\% of the total energy density of the infinite disk. The other
10\% of the energy density is distributed along the plane $z=0$ from
outside the disk to infinity. This allow us, if necessary, to treat the
outer 10\% of the energy density disk as a perturbation in the outmost
boundary condition. Moreover, the values of the parameters ($a,m$) of the
finite disk we create must satisfy the dominant energy condition. For the
infinite disk, this condition imposes that $\tilde{\sigma} + \tilde{p_r} +
\tilde{p_\varphi} >0$ for all $r$. This inequality is valid in all the
disk if $m < a$, the lower the value of $a$ for a given $m$ the faster the
surface energy density decreases. Furthermore, this condition also implies
that no tachyonic matter is present in the disk and that $p_r$ and
$p_\varphi$ are in fact pressures and not tensions.

The second order equation (\ref{2order}) is solved with two boundary
conditions, one in $r \approx 0$ and the other in the cut-off radius set
by the criterion established above. At $r \approx 0$ we set the
perturbation to be $\approx 10\%$ of the unperturbed energy density value
and at the edge of the disk $\delta \sigma |_{r=R_{cut}} = 0$. The last
condition is imposed because we want the perturbation to vanish when $r$
tends to the outer radius.

Now, we consider the isotropic Schwarzschild thin disk with parameters
($a=0.5,m=0.4$). With these parameters, the outer radius of the disk is
set to $r=4$ (approximately 90\% of the energy density is inside the
disk). In Fig.  \ref{k0k1k2} we show the amplitude profiles of the true
energy density perturbation $\tilde{\delta \sigma}$ for different modes of
the perturbation (\ref{perturbation}). We see from Fig. \ref{k0k1k2} that
the energy density perturbation profiles are stable and have an
oscillatory character.  When we increase the parameter $w$ the number of
oscillations within the disk increases and when we increase the wave
number $k$ the amplitude of the oscillations decreases. Note that the
amplitudes of the modes decay quickly when we increase the value of the
wave number. From Fig. \ref{k0k1k2} we see that there is a factor of
approximately $10^{-3}$ between the modes $(w=1,k=0)$ and $(w=1,k=2)$.

In Fig. \ref{velocity}, we present the amplitude profiles of the true
pressure perturbation and the physical radial velocity perturbation
($\tilde{\delta U^r} = \delta U^r \sqrt{g_{rr}}$) for the first three $w$
modes with $k=0$, and the amplitude profile of the physical azimuthal
velocity perturbation ($\tilde{\delta U^\varphi} = \delta U^\varphi
\sqrt{g_{\varphi\varphi}}$) for the first three $w$ modes with $k=1$.  
The modes of the azimuthal velocity perturbation with $k=0$ are equal to
zero. We see from Fig. \ref{velocity} that the pressure perturbation has
the same qualitative aspects of the density perturbation. The azimuthal
velocity perturbation amplitude shows an oscillatory behavior but the
difference between it and the radial velocity perturbation is that after
the first oscillations the maximum value of the amplitude remains almost
constant for the rest of the disk. Note that the amplitude of the radial
velocity perturbation increases with the radial coordinate. To discuss
about this fact, we have to take into account that we are using in our
numerical calculations a finite disk instead of the infinite disk exact
solution. In the infinite disk there is no place for stars to escape from
it but in the finite disk this could happen. For our disk to be
consistent, the values of the radial velocity perturbation within the disk
can not be greater than the particle escape velocity. In first
approximation, the escape velocity of our disk is calculated using
Newtonian mechanics. In the Newtonian limit of General Relativity, we have
the well known relation $g_{00} = \eta_{00} + 2 \Phi$ and the particle
escape velocity $V_e = \sqrt{2 |\Phi|}$ can be written as

\begin{equation}
V_e = \left(1 - \frac{(1-\frac{m}{2R})^2}{(1+\frac{m}{2R})^2} 
\right)^{1/2}.
\end{equation} 

\noindent For ($a=0.5,m=0.4,r=4$) the escape velocity is $\approx 0.42$
and we see from the radial velocity perturbation profile of Fig.
\ref{velocity} that the particles within the disk do not escape. In the
case when the outer radius is equal to 5 (approximately 92\% of the energy
density)  and the same parameters and perturbations are set, the escape
velocity is $\approx 0.38$ and some modes of the radial velocity
perturbation violates the consistency of the disk and can not exist. In
other words, these modes describe a disk of a certain cut-off radius with
no particles inside.

The case studied in this section corresponds to an extreme disk because
the radial velocity of the particles at the cut-off radius are near the
particle escape velocity. This example was chosen by didactic reasons to
show the problems that may appear in finding a consistent finite disk
model, but in most of the cases these problems do not appear. Recall that
the perturbation made to the energy density was 10\% of its unperturbed
initial density. If we limited our perturbation to lower values, the
radial velocity perturbation through the disk is below the escape
velocity. Furthermore, we can define the cut-off radius of the disk when
the energy density within the disk is lower than 90\%. With lower values
for the center perturbation and percentage of energy density within the
disk, all modes tested were stable.  The phenomenon of particles escaping
from the disk is more suitable to occur in the mode with $(k=0,w=1)$. We
see from Fig. \ref{k0k1k2} and Fig.  \ref{velocity} that increasing the
number of the wave vector $k$ the amplitude of the variables considered
decreases. All these considerations suggest that the isotropic
Schwarzschild thin disk is stable under a general first order perturbation
of the form (\ref{perturbation}) and favors the formation of rings.

When we performed the radial cut in the infinite disk in order to create
the finite disk, we laid outside a fraction of the total energy density
distributed up to infinity. We can model the presence of this exterior
energy density, if necessary, by perturbing the outer boundary condition,
i.e., setting in the outer radius the condition $\delta \sigma
|_{r=R_{cut}} = \epsilon$ where $\epsilon \ll \delta \sigma |_{r=0}$. The
numerical experiments realized show that this outer perturbation does not
modify the qualitative aspects of our solutions. However, depending on the
value of the perturbation, the quantitative aspects do change, but the
disk remains stable.

\section{Thin disks with only azimuthal pressure} \label{section3}

\subsection{Chazy-Curzon thin disk} \label{subsection3.1}

The Chazy-Curzon thin disk in Weyl coordinates \cite{bic:lyn} has only
azimuthal pressure, the absence of a radial pressure turns this solution
rather unphysical. Even though, one can argue that the stability of the
system can be achieved if we have two circular streams of particles moving
in opposite directions, i.e. the counter-rotating hypothesis.  
Furthermore, the dragging of inertial frame of a rotating disk does not
generate a Weyl type metric but the counter-rotating hypothesis solve this
problem. Recently, evidence of disks made of streams of rotating and
counter-rotating matter has been found \cite{ber}. Also, a detailed study
of the counter-rotating model for generic finite axially symmetric thin
disk without radial pressure can be found in Ref. \cite{gon:esp}.

The Chazy-Curzon metric is obtained from (\ref{metric}) with

\begin{eqnarray}
&&\Phi = -\frac{m}{R}, \nonumber \\
&&\Lambda = -\frac{m^2 r^2}{2 R^4}, 
\end{eqnarray}

\noindent where $R^2=r^2+(|z|+a)^2$, and

\begin{eqnarray}
&&\Psi_1 = -\Psi_2 = \Phi, \nonumber \\
&&\Psi_3 = \Lambda-\Phi. \label{cc}
\end{eqnarray}

\noindent The surface energy density and azimuthal pressure of the the
Chazy-Curzon thin disk, after applying the image method to the metric
(\ref{metric}) with the above definitions, are given by \cite{bic:lyn},

\begin{eqnarray}
&&\sigma= \frac{m a}{ 2 \pi R_0^3} \left[1 - \frac{m}{R_0} 
\left( 1 - \frac{a^2}{R_0^2} \right) \right] e^{2(\Phi_0-\Lambda_0)}, 
\label{ccsigma} \\
&&p_\varphi = \frac{m^2 a}{2 \pi R_0^4} \left( 1 
-\frac{a^2}{R_0^2} \right) e^{2(\Phi_0-\Lambda_0)}, \label{ccp}
\end{eqnarray}

\noindent where $R_0$, $\Phi_0$ and $\Lambda_0$ are the respective
functions evaluated at $z=0$. Since the Chazy-Curzon disk has no radial
pressure, we have $p_r = \delta p_r = 0$ in the perturbed equations.
Substituting $\delta U^r$ and $\delta U^\varphi$ from Eqs. (\ref{thinr})
and (\ref{thinvarphi}) into Eq. (\ref{thint}) and using a relation similar
to (\ref{psigma}) involving the fluid variables (\ref{ccsigma}) and
(\ref{ccp}), we obtain a first order differential equation in $r$ for the
perturbation $\delta \sigma$ of the form

\begin{figure}
\vspace{0.7cm}
\epsfig{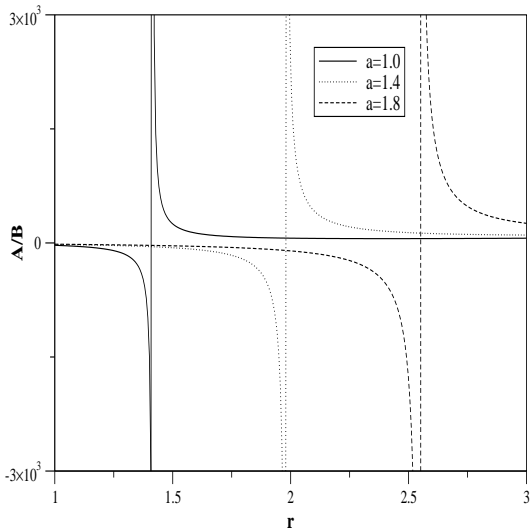} 
\caption{Profiles of the $A/B$ function present in the solution of the 
energy density perturbation for the thin Chazy-Curzon disk. The profiles 
show a singularity (instability) that depends on the cut parameter 
$a$. For $a=(1;\;1.4;\;1.8)$ and $m=1$, the singularities are 
approximately at $(1.414;\;1.980;\;2.546)$ respectively.} 
\label{ccw1k0} 
\end{figure}

\begin{equation}
A \delta \sigma_{,r} + B \delta \sigma = 0, \label{foe}
\end{equation}

\noindent where ($A,B$) are functions of ($r,a,m,w,k_\varphi$), see
Appendix B. The above equation has for solution

\begin{equation}
\delta \sigma = e^{\int(A/B){\rm d}r}. \label{sfoe}
\end{equation}

\begin{figure}
\vspace{0.7cm}
\epsfig{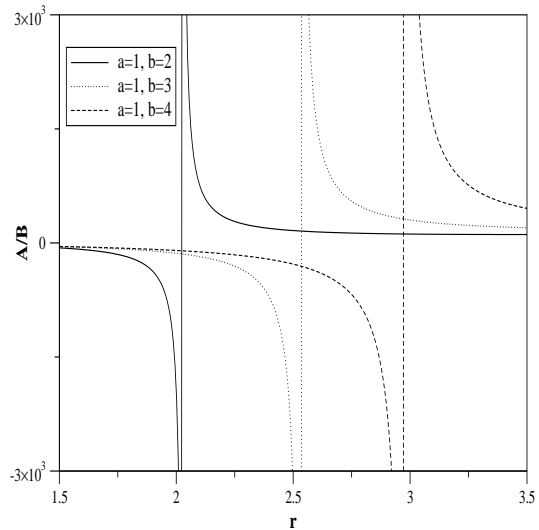} 
\caption{Profiles of the $A/B$ function present in the solution of the
energy density perturbation for the thin Zipoy-Voorhees disk. As the
Chazy-Curzon case, the profiles show a singularity (instability) that
depends on the cut parameters $(a,b)$. For $b=(2;\;3;\;4)$, $a=1$ and 
$m=1$, the singularities are approximately at $(2.193;\;2.523;\;2.978)$
respectively.}
\label{zvw1k0}
\end{figure}

\noindent In the integral argument, the values of ($a,m$) must satisfy the
dominant energy condition $\tilde{\sigma} + \tilde{p_\varphi} >0$ for all
$r$ if we want a fluid made of matter with the usual gravitational
attractive property. This inequality is satisfied if $m/a \leq 3
\sqrt{3}/4 \approx 1.30$. In the case of the Chazy-Curzon disk, the
dominant energy condition implies that the velocity of the
counter-rotating streams does not exceed the velocity of light. We show in
Fig. \ref{ccw1k0} the profile of the integral argument $A/B$ present in
the solution (\ref{sfoe}) for different value of the parameter $a$. The
singular points do not depend on the values of the parameters
($m,w,k_\varphi$)  but on the value of the cut parameter $a$. The
numerical experiments realized for the true surface energy density
$\tilde{\delta \sigma}$ show instabilities at the same values of the
singularities present in the function $A/B$. As far as we can test, the
Chazy-Curzon disk does not allow any stable mode under perturbations of
the form (\ref{perturbation}). Stability can be achieved if we let $p_r=0$
and $\delta p_r \neq 0$, but in this case is being changed the state
equation of the fluid on the disk.

\subsection{Zipoy-Voorhees thin disk} \label{subsection3.2}

In this section we study the stability of the Zipoy-Voorhees thin disks.
These disks are obtained from the Zipoy-Voorhees metric, also known as the
$\gamma$-metric. This metric represents, in Weyl coordinates, a uniform
rod of length $b-a$ at a distance $a$ below the $z=0$ plane along the
negative $z$ axis. The Zipoy-Voorhees metric is obtained from
(\ref{metric}) with

\begin{eqnarray}
&&\Phi = \frac{m}{b-a} \ln \left[ \frac{R_a + |z| + a}{R_b + |z| + b} 
\right], \nonumber \\
&&\Lambda = \frac{2 m^2}{(b-a)^2} \ln \left[ \frac{(R_a+R_b)^2-(b-a)^2}{4 
R_a R_b} \right],
\end{eqnarray}

\noindent where $R_a^2=r^2+(|z|+a)^2$ and $R_b^2=r^2+(|z|+b)^2$, and

\begin{eqnarray}
&&\Psi_1 = -\Psi_2 = \Phi, \nonumber \\
&&\Psi_3 = \Lambda-\Phi.
\end{eqnarray}

\noindent When $(b-a) = m$, the above solution leads to the Schwarzschild
metric and when $(b-a) \rightarrow 0$, to the Chazy-Curzon solution. The
thin disk formed by applying the displace, cut and reflect method has the
surface energy density and the azimuthal pressure given by \cite{bic:lyn},

\begin{eqnarray}
&&\sigma= \frac{m}{2 \pi (b-a)} \left(\frac{1}{R_{a0}} - 
\frac{1}{R_{b0}} \right) \\
&& \hspace{1.5cm} \left[1 - \frac{m}{b-a} \left( \frac{b}{R_{b0}} - 
\frac{a}{R_{a0}} \right) \right] e^{2(\Phi_0 - \Lambda_0)}, \nonumber \\
&&p_\varphi= \frac{m^2}{2 \pi (b-a)^2} \left( \frac{1}{R_{a0}} - 
\frac{1}{R_{b0}} \right) \\ 
&& \hspace{1.5cm} \left( \frac{b}{R_{b0}} -
\frac{a}{R_{a0}} \right) e^{2(\Phi_0 - \Lambda_0)}, \nonumber
\end{eqnarray}

\noindent where $b \geq a$ and $R_{a0}$, $R_{b0}$, $\Phi_0$ and
$\Lambda_0$ are the respective quantities evaluated at $z=0$. Like the
Chazy-Curzon metric, these disks do not have radial pressure and, in order
to generate a Weyl type metric, the counter-rotating hypothesis is 
applied.

Following the same steps as in the Chazy-Curzon metric, we arrive to a
first order differential equation that yields a solution of the form
(\ref{sfoe}). In this case, the values of the parameters ($a,b,m$) must
satisfy the dominant energy condition $\tilde{\sigma} + \tilde{p_\varphi}
>0$. To achieve this condition we must have \cite{bic:lyn},

\begin{equation}
\frac{m}{b-a} \leq \frac{1}{2} \left( \frac{R_{b0} R_{a0}}{b R_{a0} - a 
R_{b0}} \right),
\end{equation}

\noindent when 

\begin{equation}
r^2 = b^2 \left( \frac{a}{b} \right)^{2/3} \left[1+ \left( \frac{a}{b} 
\right)^{2/3} \right].
\end{equation}

\noindent In Fig. \ref{zvw1k0} we plot the profiles of the integral
argument $A/B$ for the Zipoy-Voorhees disk for different values of the
parameter $b$. As the Chazy-Curzon disks, the singular points that appear
in the profiles do not depend on the values of ($m,w,k_\varphi$)  but on
the value of the cut parameters $(a,b)$. Also, numerical experiments show
that the true surface energy density is not stable at the same values of the
singularities present in the function $A/B$. Moreover, under perturbation
of the form (\ref{perturbation}), the Zipoy-Voorhees disk does not have
any stable mode.  A different treatment of the stability of the
Chazy-Curzon and Zipoy-Voorhees thin disks \cite{bic:lyn}, made by
analyzing their velocity curves and specific angular momentum, shows that
they are not stable for highly relativistic disks.

\section{Conclusions} \label{section5}

We study the stability of several thin disk models in the context of
General Relativity using a general first order perturbation. The stability
analysis performed in this work is better than the stability analysis of
particle motion along geodesics because we take into account the
collective behavior of the particles. However, this analysis can be said
to be incomplete because the energy momentum perturbation is treated like
a test fluid and does not alter the background metric. The complete
analysis of the thin disks stability needs to take into account the
perturbation of the metric. This is a second degree of approximation to
the stability problem in which the emission of gravitational radiation is
considered.

We show that the Chazy-Curzon and Zipoy-Voorhees thin disks are not stable
under perturbations of the form (\ref{perturbation}). This is due to the
fact that there is no radial pressure to support the gravitational
attraction of the disk. This appears to be in contradiction with the
experimental evidence of stellar systems made of streams of rotating and
counter-rotating matter \cite{ber}, but the counter-rotating hypothesis
assumes the existence of two pressureless streams of matter in circular
opposite orbits which do not interact, see appendix of Ref.  \cite{let2},
and this is not the case.

In the case of the isotropic Schwarzschild thin disk, where we have radial
and azimuthal pressures, we formed a finite disk by making a cut-off
radius and allowing a percentage of the unperturbed energy density within
the disk. The finiteness of our new disk allows the particles to escape
from it, so we have to compare the particles escape velocity with the
velocity perturbation profiles if we want to have a self-consistent finite
thin disk model or if we want to discard some perturbation modes. If we
lower the value of the initial perturbation and/or the percentage of the
total energy density inside the disk, all the modes are stable. The fluid
variables, in the isotropic Schwarzschild thin disk, present an
oscillatory character with the amplitudes vanishing when $r$ approaches
the outmost radius. In the case of the azimuthal perturbation, the
amplitude is almost constant within the disk. In general, when we increase
the parameter $w$ the number of oscillations increases inside the disk and
the amplitudes decrease.  When we increase the wave number $k$ the values
of the amplitudes decrease abruptly. We note that the perturbation
(\ref{perturbation}) made in the isotropic Schwarzschild thin disk favors
the formation of rings. As expected, the presence of a radial pressure is
fundamental to the stability of thin disks.

\section*{ACKNOWLEDGMENTS}

M.U. and P.S.L. thanks FAPESP for financial support and  P.S.L. also thanks
CNPq.

\section*{APPENDIX A: COEFFICIENTS OF THE ISOTROPIC SCHWARZSCHILD THIN DISK
SECOND ORDER PERTURBATION  EQUATION}

The general form of the functions $(A_S,B_S,C_S)$ appearing in the second
order equation (\ref{2order}) are given by

\begin{eqnarray}
&&A_S = A_1 \alpha_1, \nonumber \\
&&B_S = A_1 (\alpha_{1,r} + \alpha_2) + A_2 \alpha_1, \nonumber \\
&&C_S = A_1 \alpha_{2,r} + A_2 \alpha_2 + A_3 \alpha_3 + A_4, \label{a1} 
\end{eqnarray}

\noindent where $\alpha_1$, $\alpha_2$ and $\alpha_3$ are

\begin{eqnarray}
&&\alpha_1 = -\frac{f_{,r} B_1 + f (B_4 + B_5) + B_3}{B_2}, \nonumber \\
&&\alpha_2 = -\frac{f B_1}{B_2}, \nonumber \\
&&\alpha_3 = -\frac{f C_2}{C_1}, \label{a2}
\end{eqnarray}

\noindent and $f=(p_{,r}/\sigma_{,r})$. In Eqs. (\ref{a1}) and (\ref{a2}),
we denote the coefficients of Eq.  (\ref{thint}) by $A_i$, the
coefficients of Eq. (\ref{thinr}) by $B_i$ and the coefficients of Eq.
(\ref{thinvarphi}) by $C_i$, e.g., the first term in (\ref{thint}) has the
coefficient $A_1$ multiplied by the factor $\delta U^r_{,r}$, the second
term has the coefficient $A_2$ multiplied by the factor $\delta U^r$, etc.

The explicit form of the above equations are obtained substituting the
fluid variables ($p,\sigma$) of the isotropic Schwarzschild thin disk.

\section*{APPENDIX B:  ARGUMENT $(A/B)$ OF THE INTEGRAL  (\ref{sfoe})}

The general form of the  argument  $(A/B)$  of the integral 
Eq. (\ref{sfoe}) are given by

\begin{eqnarray}
&& A = \sigma U^t \alpha_{1,r} + {\rm F}(t,r,\sigma U^t) \alpha_1
\nonumber \\
&&\hspace{0.8cm} - k_\varphi (\sigma U^t + \xi_2 p_\varphi Y^\varphi ) 
\alpha_2 + w U^t U^t, \nonumber \\
&&B = \sigma U^t \alpha_1,
\end{eqnarray}

\noindent where the functions $\alpha_1$ and $\alpha_2$ are

\begin{eqnarray}
&&\alpha_1 = \frac{ U^t U^t \Gamma^r_{tt} + Y^\varphi Y^\varphi 
\Gamma^r_{\varphi\varphi} (p_{\varphi,r}/\sigma_{,r})}{w\sigma U^t}  \\
&&\alpha_2 = \frac{k_\varphi Y^\varphi Y^\varphi}{w(\sigma U^t + \xi_2 
p_\varphi Y^\varphi)}.
\end{eqnarray}

\noindent The explicit form of the above equations are obtained
substituting the particular fluid variables ($p_\varphi,\sigma$) of 
the Chazy-Curzon or Zipoy-Voorhees thin disks.

\end{document}